\documentclass[11pt]{article}
\usepackage{pgf,tikz}
\usetikzlibrary{calc,arrows}
\usepackage[all]{xy}
\usepackage{amsmath}
\usepackage{amsfonts}
\usepackage{amssymb}
\usepackage{latexsym}
\usepackage{graphicx}
\usepackage{color}
\usepackage{amscd}
\usepackage{epsfig}
\usepackage{cite}
\usepackage{circuitikz}
\usepackage[perpage,symbol]{footmisc}
\usepackage{listings}
\usepackage{array}

\newtheorem{problem}{Problem}[section]

\newtheorem{lemma}[problem]{Lemma}
\newtheorem{theorem}[problem]{Theorem}

\newtheorem{corollary}[problem]{Corollary}

\title{Subadditive stake systems}
\author{Chunlei Liu\footnote{Shanghai Dengbi Comm. Tech. Co. Ltd., Shanghai, China. 714232747@qq.com}}
\date{}
\begin{document}
\maketitle
\abstract{We propose stake system which issues stakes as well as coins. Cash systems can be viewed as stake systems. Two subadditive stake systems are studied: the radical stake system and the logarithmic stake system.  In both subadditive stake systems, an attacker would find it very difficult to build the longest block-chain alone.}

\section{\small{INTRODUCTION}}
\paragraph{}
In 2009, Satoshi Nakamoto \cite{Na} introduced the notion of block-chain into P2P cash systems, giving birth to the famous Bitcoin, which is the first P2P cash implemented  in practise.
\paragraph{}
A cash system is a system which issues coins, and in which nodes transfer coins to each other. A P2P cash system is a cash system with a digital signature scheme in which transactions are digitally signed and are broadcast to all nodes. A block-chain cash system with a hash function $B\mapsto{\rm hash}(B)$ and a threshold function $B\mapsto{\rm threshold}(B)$ is a P2P cash system, where transactions are collected into blocks,
 where the hash of a block is contained in the next block so that the blocks are chained one after another, where only the longest block-chain is considered to correct,
 where a nonce is added to a block so that $${\rm hash}(B)\leq {\rm threshold}(B),\forall B,$$
and where an amount of ${\rm Rwd}$ new coins are rewarded to a block creator.
\paragraph{}A block-chain cash system is said to be based on proof of work if
$${\rm threshold}(B)=\frac{M}{D},\ \forall B,$$
where $M$ is the scale of the system, and $D$ is the difficulty constant of the system.
\paragraph{}A block-chain cash system is said to be based on proof of stake if
$${\rm threshold}(B)=M\cdot\frac{{\rm bal}(A;C)+{\rm Rwd}}{D},\ \forall B,$$
where $M$ is the scale of the system, $D$ is the difficulty constant of the system, $A$ is the creator of $B$, $C$ is the block-chain after which $B$ is chained, ${\rm bal}(A;C)$ is the balance of $A$ in $C$, and ${\rm Rwd}$ is the amount of new coins awarded to a block creator.
\paragraph{}The block-chain cash system based on proof of stake have been studied by many authors \cite{KN, BGM, NXT, Mi, BPS, DGKR, KRDO}. However,
the block-chain cash system based on proof of stake seems vulnerable to long-term attacks, see, e.g. \cite{Bu, Po}.
\paragraph{}We now propose stake systems.
A stake system is a cash system which issues stakes as well as coins, in which nodes transfer coins to each other, and in which transaction fees are paid with coins. A P2P stake system is a stake system with a digital signature scheme in which transactions are digitally signed and are broadcast to all nodes. A block-chain stake system with a hash function $B\mapsto{\rm hash}(B)$, a coin-issue threshold function $B\mapsto{\rm threshold}(B)$, and a stake-issue threshold function $B\mapsto{\rm stakthreshold}(B)$ which is majored by the coin-issue threshold function is a P2P stake system where transactions are collected into blocks,
 where the hash of a block is contained in the next block so that the blocks are chained one after another, where only the longest block-chain is considered to correct,
 where a nonce is added to a block so that $${\rm hash}(B)\leq {\rm cointhreshold}(B),\forall B,$$
 where an amount of ${\rm CoinRwd}$ new coins are rewarded to a block creator, and where an amount of ${\rm StakRwd}$ new stakes are rewarded to a block creator if he has created a block, say $B$, which satisfies $${\rm hash}(B)\leq {\rm stakthreshold}(B).$$
\paragraph{} A block-chain cash system may be regarded as a block-chain stake system whose stake-issue threshold is the same as its coin-issue threshold, and in which stakes are never transferred to each other so that the stakes of a node is just the product of ${\rm Rwd}$ and the times he has got rewarded.
\paragraph{} A block-chain cash system may also be regarded as a block-chain stake system whose stake-issue threshold is the same as its coin-issue threshold, and in which coins ever used to pay transaction fees lost their stakes so that the stakes of a node is the sum of the part of coins he owned but is never used to pay transaction fees and the part of transaction fees he has paid with coins  which is used to pay transaction fees for the first time.
\section{\small{CONSTANT STAKE SYSTEMS}}
\paragraph{}A block-chain stake system is called a constant stake system if if
$${\rm cointhreshold}(B)=\frac{M}{{\rm CoinD}},\ \forall B,$$
and
$${\rm stakthreshold}(B)=\frac{M}{{\rm StakD}},\ \forall B,$$
where $M$ is the scale of the system, ${\rm CoinD}$ and ${\rm StakD}$ are respectively the coin-issue difficulty constant and the stake-issue difficulty constant of the system.

\paragraph{}The block-chain cash system based on proof of work may be regarded as a constant stake system in which ${\rm CoinD}={\rm StakD}$. It is easy to see that a constant stake system is as secure as the block-chain cash system based on proof of work.
\section{\small{LINEAR STAKE SYSTEMS}}
\paragraph{}A block-chain stake system is called a linear stake system if
$${\rm cointhreshold}(B)=M\cdot\frac{{\rm stak}(A;C)+{\rm StakRwd}}{{\rm CoinD}},\ \forall B,$$
and
$${\rm stakthreshold}(B)=M\cdot\frac{{\rm stak}(A;C)+{\rm StakRwd}}{{\rm StakD}},\ \forall B,$$
where $M$ is the scale of the system, ${\rm CoinD}$ and ${\rm StakD}$ are respectively the coin-issue difficulty constant and the stake-issue difficulty constant of the system, $A$ is the creator of $B$, $C$ is the block-chain after which $B$ is chained, ${\rm stak}(A;C)$ is the stake of $A$ in $C$, and ${\rm StakRwd}$ is the amount of new stakes awarded to a block-creator when the hash of the created block is no greater than the stake-issue threshold.
\paragraph{}
Though a linear stake system is a little different from a block-chain cash system based on proof of stake, it is still not resistant to long-term attacks.
\section{\small{RADICAL STAKE SYSTEMS}}
\paragraph{}Let $0<a<1$. A stake system is called a radical stake system with equal exponent $a$ if
$${\rm cointhreshold}(B)=M\cdot\frac{({\rm StakRwd}+{\rm stak}(A,C))^a}{\rm CoinD}$$
and
$${\rm stakthreshold}(B)=M\cdot\frac{({\rm StakRwd}+{\rm stak}(A,C))^a}{\rm StakD},$$
where $M$ is the scale of the system, ${\rm CoinD}$ and ${\rm StakD}$ are respectively the coin-issue difficulty constant and the stake-issue difficulty constant of the system, $A$ is the creator of $B$, $C$ is the block-chain after which $B$ is chained, ${\rm stak}(A;C)$ is the stake of $A$ in $C$, and ${\rm StakRwd}$ is the amount of new stakes awarded to a block-creator when the hash of the created block is no greater than the stake-issue threshold.
We now prove the following.
\begin{theorem}Suppose that a node, who conducts no transactions on stakes with other nodes, is going to build a block-chain alone. Then the expected time for the party to build a block-chain of length $L$ in a radical stake system with equal exponent $a$ is
$$\frac{\rm CoinD}{{\rm StakRwd}^a}\sum_{n=0}^{L-1}\sum_{k=0}^n\frac{1}{(k+1)^a}{n\choose k}p^kq^{n-k},$$
 where $p=\frac{\rm CoinD}{\rm StakD}$, and $q=1-p$.
\end{theorem}
{\it Proof.} Note that, after the node has built
$n$ blocks, the probability for him to be rewarded with stakes $k$ times is ${n\choose k}p^kq^{n-k}$.
 So the expected time for the node to chain the $(n+1)$-th block is
$$\sum_{k=0}^n\frac{\rm CoinD}{(k+1)^a\cdot{\rm StakRwd}^a}{n\choose k}p^kq^{n-k}$$
$$=\frac{\rm CoinD}{{\rm StakRwd}^a}\sum_{k=0}^n\frac{1}{(k+1)^a}{n\choose k}p^kq^{n-k}.$$
 It follows that the expected time for the node to build a long  block-chain of length $L$ is
$$\frac{\rm CoinD}{{\rm StakRwd}^a}\sum_{n=0}^{L-1}\sum_{k=0}^n\frac{1}{(k+1)^a}{n\choose k}p^kq^{n-k}.$$
The theorem is proved.
\begin{corollary}Suppose that a node, who conducts no transactions on stakes with other nodes, is going to build a block-chain alone. Then the expected time for the party to build a block-chain of length $L$ in a radical stake system with equal exponent $a$ in which ${\rm CoinD}={\rm StakD}$ is
$$\frac{\rm CoinD}{{\rm StakRwd}^a}\sum_{n=0}^{L-1}\frac{1}{(n+1)^a}.$$
\end{corollary}
{\it Proof.} As ${\rm CoinD}={\rm StakD}$, we have $p=1$ and $q=0$, and hence
$$\sum_{k=0}^n\frac{1}{(k+1)^a}{n\choose k}p^kq^{n-k}=\frac{1}{(n+1)^a}.$$
The corollary now follows.
\begin{lemma}We have
$$\sum_{k=0}^n\frac{1}{(k+1)^a}{n\choose k}p^kq^{n-k}\leq\frac1p\cdot\frac{1-q^{n+1}}{(n+1)^a}.$$
\end{lemma}
{\it Proof.} We have
$$\sum_{k=0}^n\frac{1}{(k+1)^a}{n\choose k}p^kq^{n-k}$$
$$=\frac{1}{(n+1)^a}\sum_{k=0}^n\frac{(n+1)^a}{(k+1)^a}{n\choose k}p^kq^{n-k}$$
$$\leq\frac{1}{(n+1)^a}\sum_{k=0}^n\frac{n+1}{k+1}{n\choose k}p^kq^{n-k}$$
$$\leq\frac1p\cdot\frac{1-q^{n+1}}{(n+1)^a}.$$
The lemma is proved.
\begin{corollary}Suppose that a node, who conducts no transactions on stakes with other nodes, is going to build a block-chain alone. Then the expected time for the party to build a block-chain of length $L$ in a radical stake system with equal exponent $a$ is no greater than
$$\frac{\rm StakD}{{\rm StakRwd}^a}\sum_{n=0}^{L-1}\frac{1}{(n+1)^a}.$$
\end{corollary}
The above corollary says that a node, who conducts no transactions on stakes with other nodes and is going to build a block-chain alone, gets no faster if he doesn't add  a new block to the block-chain until the hash of the block is no greater than the stake-issue threshold.
\begin{theorem}\label{chaintime}Suppose that a party with $m\geq2$ nodes is going to build a block-chain. Assume that the party conducts no transactions on stakes with nodes outside the party. Let $X_i$ be the proportion of stakes of the $i$-th node.  Then the expected time for the party to build a block-chain of length $L$ in a radical stake system with equal exponent $a$ is no greater than
$$E((\sum_{i=1}^mX_i^a)^{-1})\frac{{\rm CoinD}}{{\rm StakRwd}^a}\sum_{n=1}^L\sum_{k=0}^n\frac{1}{(k+1)^a}{n\choose k}p^kq^{n-k},$$
 where $p=\frac{\rm CoinD}{\rm StakD}$, $q=1-p$, and $E((\sum_{i=1}^mX_i^a)^{-1})$ is the expectation of $(\sum_{i=1}^mX_i^a)^{-1}$.
\end{theorem}
{\it Proof.} Note that, after the party has built
$n$ blocks, the probability that the party is rewarded with stakes $k$ times is ${n\choose k}p^kq^{n-k}$.
Let $T$ be the time for the party creates the $(n+1)$-th block with the unit time being the time for a CPU to perform one operation.
Then
$$P(T=t)=\sum_{k=0}^n{n\choose k}p^kq^{n-k}\sum_{\vec{x}}f(\vec{x})(\alpha(\vec{x})^{t-1}-\alpha(\vec{x})^t),
$$
where $f(\vec{x})$ is the probability mass function of the random variable $(X_1,\cdots,X_m)$, and $$\alpha(\vec{x})=\prod_{i=1}^m(1-\frac{(1+k x_i)^a{\rm StakRwd}^a}{\rm CoinD}).$$
So the expected time for the node to chain the $(n+1)$-th block is
$$\sum_{k=0}^n{n\choose k}p^kq^{n-k}\sum_{\vec{x}}f(\vec{x})
\frac{\rm CoinD}{\sum_{i=1}^m(kx_i+1)^a\cdot{\rm StakRwd}^a}$$
Note that $$kx+1\geq (k+1)x,\ 0\leq x\leq1.$$
So the expected time for the node to chain the $(n+1)$-th block is no greater than $$
\leq\frac{\rm CoinD}{{\rm StakRwd}^a}\sum_{k=0}^n\frac{1}{(k+1)^a}{n\choose k}p^kq^{n-k}\sum_{\vec{x}}
\frac{f(\vec{x})}{\sum_{i=1}^mx_i^a}.$$
The theorem is proved.
\paragraph{}Note that
$$E((\sum_{i=1}^mX_i^a)^{-1})<1.$$
Therefore by the above theorems, it is very difficult for an attacker to build the longest block-chain alone.
To get a sense of the degree of the difficulty an attacker would face when he started to build the longest chain, we prove the following lemma.
\begin{lemma}Let $X_i$ be the proportion of stakes of the $i$-th node in a party with $m$ nodes. Let $c>1$. Suppose that
the probability mass function of $(X_1,\cdots,X_m)$ vanishes at all points $(x_1,\cdots,x_m)$ for which
$$|\{i\mid x_i>\frac{1}{m}\}|<cm^a.$$ Then
$$E((\sum_{i=1}^mX_i^a)^{-1})\leq\frac1c.$$\end{lemma}
{\it Proof. }Note that, $$\sum_{i=1}^mx_i^a\geq c\text{ whenever }|\{i\mid x_i>\frac{1}{m}\}|\geq cm^a.$$
So
$$E((\sum_{i=1}^mX_i^a)^{-1})\leq E(\frac1c)\leq\frac1c.$$
The lemma is proved.

\section{\small{LOGARITHMIC STAKE SYSTEMS}}
\paragraph{}A stake system is called a logarithmic stake system if
$${\rm cointhreshold}(B)=M\cdot\frac{\log_2({\rm StakRwd}+{\rm stak}(A,C))}{\rm CoinD}$$
and
$${\rm stakthreshold}(B)=M\cdot\frac{\log_2({\rm StakRwd}+{\rm stak}(A,C))}{\rm StakD},$$
where $M$ is the scale of the system, ${\rm CoinD}$ and ${\rm StakD}$ are respectively the coin-issue difficulty constant and the stake-issue difficulty constant of the system, $A$ is the creator of $B$, $C$ is the block-chain after which $B$ is chained, ${\rm stak}(A;C)$ is the stake of $A$ in $C$, and ${\rm StakRwd}$ is the amount of new stakes awarded to a block-creator when the hash of the created block is no greater than the stake-issue threshold.
We now prove the following.
\begin{theorem}Suppose that a node, who conducts no transactions on stakes with other nodes, is going to build a block-chain alone. Then the expected time for the party to build a block-chain of length $L$ in a logarithmic stake system is
$$={\rm CoinD}\sum_{n=0}^{L-1}\sum_{k=0}^n\frac{1}{\log_2(k+1)+\log_2{\rm StakRwd}}{n\choose k}p^kq^{n-k},$$
 where $p=\frac{\rm CoinD}{\rm StakD}$, and $q=1-p$.
\end{theorem}
{\it Proof.} Note that, after the node has built
$n$ blocks, the probability for him to be rewarded with stakes $k$ times is ${n\choose k}p^kq^{n-k}$.
 So the expected time for the node to chain the $(n+1)$-th block is
$$\sum_{k=0}^n\frac{\rm CoinD}{\log_2(k+1)+\log_2{\rm StakRwd}}{n\choose k}p^kq^{n-k}$$
$$={\rm CoinD}\sum_{k=0}^n\frac{1}{\log_2(k+1)+\log_2{\rm StakRwd}}{n\choose k}p^kq^{n-k}.$$
 It follows that the expected time for the node to build a long  block-chain of length $L$ is
$$={\rm CoinD}\sum_{n=0}^{L-1}\sum_{k=0}^n\frac{1}{\log_2(k+1)+\log_2{\rm StakRwd}}{n\choose k}p^kq^{n-k}.$$
The theorem is proved.
\begin{corollary}Suppose that a node, who conducts no transactions on stakes with other nodes, is going to build a block-chain alone. Then the expected time for the party to build a block-chain of length $L$ in a logarithmic stake system in which ${\rm CoinD}={\rm StakD}$ is
$${\rm CoinD}\sum_{n=0}^{L-1}\frac{1}{\log_2(n+1)+\log_2{\rm StakRwd}}.$$
\end{corollary}
{\it Proof.} As ${\rm CoinD}={\rm StakD}$, we have $p=1$ and $q=0$, and hence
$$\sum_{k=0}^n\frac{1}{\log_2(k+1)+\log_2{\rm StakRwd}}{n\choose k}p^kq^{n-k}$$$$=\frac{1}{\log_2(n+1)+\log_2{\rm StakRwd}}.$$
The corollary now follows.
\begin{lemma}We have
$$\sum_{k=0}^n\frac{1}{\log_2(k+1)+\log_2{\rm StakRwd}}{n\choose k}p^kq^{n-k}$$$$\leq\frac1p\cdot\frac{1-q^{n+1}}{\log_2(n+1)+\log_2{\rm StakRwd}}.$$
\end{lemma}
{\it Proof.} Note that $$\frac{\log_2(n+1)+\log_2{\rm StakRwd}}{\log_2(k+1)+\log_2{\rm StakRwd}}
\leq\frac{n+1}{k+1}.$$
So
$$\sum_{k=0}^n\frac{1}{\log_2(k+1)+\log_2{\rm StakRwd}}{n\choose k}p^kq^{n-k}$$
$$\leq\frac{1}{\log_2(n+1)+\log_2{\rm StakRwd}}\sum_{k=0}^n\frac{n+1}{k+1}{n\choose k}p^kq^{n-k}$$
$$\leq\frac1p\cdot\frac{1-q^{n+1}}{\log_2(n+1)+\log_2{\rm StakRwd}}.$$
The lemma is proved.
\begin{corollary}Suppose that a node, who conducts no transactions on stakes with other nodes, is going to build a block-chain alone. Then the expected time for the party to build a block-chain of length $L$ in a logarithmic stake system is no greater than
$${\rm StakD}\sum_{n=0}^{L-1}\frac{1}{\log_2(n+1)+\log_2{\rm StakRwd}}.$$
\end{corollary}
The above corollary says that a node, who conducts no transactions on stakes with other nodes and is going to build a block-chain alone, gets no faster if he doesn't add  a new block to the block-chain until the hash of the block is no greater than the stake-issue threshold.
\begin{theorem}\label{logchaintime}Suppose that a party with $m\geq2$ nodes is going to build a block-chain. Assume that the party conducts no transactions on stakes with nodes outside the party. Let $X_i$ be the proportion of stakes of the $i$-th node.  Then the expected time for the party to build a block-chain of length $L$ in a logarithmic stake system is no greater than
$${\rm CoinD}\cdot E((\sum_{i=1}^m\log_2(1+X_i))^{-1})\sum_{n=1}^L\sum_{k=0}^n\frac{{n\choose k}p^kq^{n-k}}{\log_2(k+1)+\log_2{\rm StakRwd}},$$
 where $p=\frac{\rm CoinD}{\rm StakD}$, $q=1-p$, and $E((\sum_{i=1}^m\log_2(1+X_i))^{-1})$ is the expectation of $(\sum_{i=1}^m\log_2(1+X_i))^{-1}$.
\end{theorem}
{\it Proof.} Note that, after the party has built
$n$ blocks, the probability that the party is rewarded with stakes $k$ times is ${n\choose k}p^kq^{n-k}$.
Let $T$ be the time for the party creates the $(n+1)$-th block with the unit time being the time for a CPU to perform one operation.
Then
$$P(T=t)=\sum_{k=0}^n{n\choose k}p^kq^{n-k}\sum_{\vec{x}}f(\vec{x})(\alpha(\vec{x})^{t-1}-\alpha(\vec{x})^{t-1}),
$$
where $f(\vec{x})$ is the probability mass function of the random variable $(X_1,\cdots,X_m)$, and $$\alpha(\vec{x})=\prod_{i=1}^m(1-\frac{\log_2(1+k x_i)+\log_2{\rm StakRwd}}{\rm CoinD}).$$
So the expected time for the node to chain the $(n+1)$-th block is
$$\sum_{k=0}^n{n\choose k}p^kq^{n-k}\sum_{\vec{x}}f(\vec{x})
\frac{\rm CoinD}{\sum_{i=1}^m(\log_2{\rm StakRwd}+\log_2(kx_i+1))}.$$
Note that
$$\log_2(1+kx)\geq\log_2(1+k)\times\log_2(1+x),\ 0\leq x\leq 1.$$ 
So the expected time for the node to chain the $(n+1)$-th block is no greater than $$
{\rm CoinD}\sum_{k=0}^n\frac{{n\choose k}p^kq^{n-k}}{\log_2(k+1)+\log_2{\rm StakRwd}}\sum_{\vec{x}}
\frac{f(\vec{x})}{\sum_{i=1}^m\log_2(1+x_i)}.$$
The theorem is proved.
\paragraph{}Note that
$$E((\sum_{i=1}^m\log_2(1+X_i))^{-1})<1.$$
Therefore by the above theorems, it is very difficult for an attacker to build the longest block-chain alone.
To get a sense of the degree of the difficulty an attacker would face when he started to build the longest chain, we prove the following lemma.
\begin{lemma}Suppose that a party with $m>2$ nodes is going to build a block-chain. Assume that the party conducts no transactions with nodes outside the party. Let $X_i$ be the proportion of stakes of the $i$-th node. Let $1<c\leq \frac{m}{\log_2(m+1)}$. Suppose that the probability mass function of $(X_1,\cdots,X_n)$ vanishes at all points $(x_1,\cdots,x_m)$ for which
$$|\{i\mid x_i>\frac{1}{m}\}|<2c\log_2m.$$ Then the expected time for the party to build a long block-chain of length $L$  is no greater than
$$\frac{\rm CoinD}{c}\sum_{n=1}^L\sum_{k=0}^n\frac{{n\choose k}p^kq^{n-k}}{\log_2(k+1)+\log_2{\rm StakRwd}},$$
 where $p=\frac{\rm CoinD}{\rm StakD}$, and $q=1-p$.\end{lemma}
{\it Proof. }We claim that, if
$$|\{i\mid x_i>\frac{1}{m}\}|\geq2c\log_2m,$$ then
$$\sum_{i=1}^m(\log_2{\rm StakRwd}+\log_2(kx_i+1))\geq c(\log_2{\rm StakRwd}+\log_2(k+1)).$$
First, if $k\leq m$, then
$$\sum_{i=1}^m(\log_2{\rm StakRwd}+\log_2(kx_i+1))$$
$$\geq m\log_2{\rm StakRwd}$$
$$\geq c(\log_2{\rm StakRwd}+\log_2(k+1)).$$
Secondly, if $k>m$, then
$$\sum_{i=1}^m(\log_2{\rm StakRwd}+\log_2(kx_i+1))$$
$$\geq m\log_2{\rm StakRwd}+2c(\log_2m)(\log_2(k+m)-\log_2m)$$
$$\geq c(\log_2{\rm StakRwd}+\log_2(k+1)).$$
The lemma now follows from the proof of Theorem \ref{logchaintime}.
\section{{\small CONCLUSION}}We have proposed stake system which issues stakes as well as coins. Two subadditive stake systems are studied: the radical stake system and the logarithmic stake system.  In both subadditive stake systems, an attacker would find it very difficult to build the longest block-chain alone.

\end{document}